\documentclass[10pt]{article}
\usepackage{hyperref,graphicx,bchart}
\usepackage[affil-it]{authblk}
\usepackage{titlesec}

\usepackage[letterpaper,margin=1.0in]{geometry}

\newcommand{\bn}{\begin{enumerate}}
\newcommand{\en}{\end{enumerate}}
\newcommand{\bi}{\begin{itemize}}
\newcommand{\ei}{\end{itemize}}

\titleformat{\section}{\large\bfseries}{\thesection}{1em}{}

\title{\vspace*{-2cm} {\large \textbf{Yet Another Snapshot of Foundational Attitudes \\ 
Toward Quantum Mechanics}}} 

\author[1]{Travis Norsen}

\author[2]{Sarah Nelson}

\affil[1]{tnorsen@smith.edu}

\affil[2]{snelson@hms.harvard.edu}

\date{June 18, 2013}

\begin{document}

\maketitle

\vspace*{-.7cm}

\begin{abstract}
\noindent  A survey probing respondents' views on various foundational
issues in quantum mechanics was recently created by Schlosshauer,
Kofler, and Zeilinger and then given to 33 participants at a quantum
foundations conference.  Here we report the results of giving this
same survey to the attendees at another recent quantum foundations
conference.  While it is rather difficult to conclude anything of
scientific significance from the poll, the results do strongly suggest
several interesting cultural facts -- for example, that there exist,
within the broad field of ``quantum foundations'', sub-communities with quite different views, and that (relatedly) there is probably even significantly more controversy about several fundamental issues than the already-significant amount revealed in the earlier poll.

\end{abstract}

\section{Introduction}

The ``snapshot of foundational attitudes toward quantum mechanics''
taken by Schlosshauer, Kofler, and Zeilinger (SKZ) and shared in
Ref.\cite{SKZ} attracted a surprising amount of attention in both the
scientific and popular media. \cite{ball,moskowitz,plumer,mit,siegfried,schloss}  Apparently many people regard it as interesting and perhaps surprising that simple, seemingly elementary questions about the meaning and implications of the (now-almost-100-years-old) quantum theory could remain unresolved and indeed hotly contested.  Actually, though, our feeling upon examining the results was that SKZ's survey did not even come close to revealing the true nature and extent of the controversy surrounding certain key issues.  

So with the hope, not so much of finding a ``truly representative
sample'' but rather simply of demonstrating the existence (and
prevalence) of different viewpoints, not well represented in the
original survey, we arranged to give the same poll to the attendees at
another recent quantum foundations conference:  ``Quantum Theory
Without Observers III'' held in Bielefeld, Germany in late April
(2013). \cite{QTWO3}  It was decided that, although a number of the
questions (and/or answers) from SKZ's poll seemed less than ideal, it
would be better to pose exactly the same set of questions so that the
answer-statistics could at least be meaningfully compared between the
two conferences.  As was done by SKZ in the original poll, the
attendees were told that their participation was optional and also
that they need not necessarily pick just one and only one answer on
each question:  multiple answers as well as write-ins were allowed.
Seventy-six people (of the roughly 100 who were in attendance) filled
out the survey in Bielefeld.

In the following section we present the results of the survey without
commentary.  A few brief thoughts are then elaborated in the
subsequent section, focusing especially on the points of most significant difference between our results and those of SKZ.

\section{\label{sec:results}Results}

For each question, the bar graph indicates the percentage of
respondents who endorsed the given options.  Afterwards we indicate
the fraction of respondents who made comments in the margin (which is
perhaps some kind of measure of how problematic people found the
question and/or the proposed answers to be) and quote any particularly
interesting or noteworthy or common comments and write-ins.  

\vspace{.2in}

\subsubsection*{Question 1: What is your opinion about the randomness of individual quantum events (such as the decay of a radioactive atom)?}

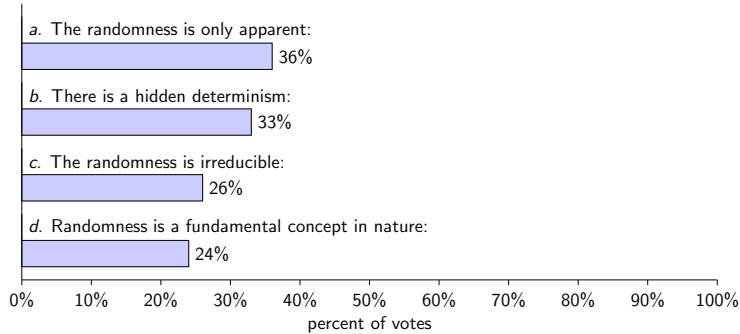
\scalebox{0.7}{\begin{bchart}[step=10,max=100,unit=\%,width=0.8\textwidth]

\bcbar[value=,text={\emph{a.} The randomness is only apparent:}]{0}
\bcbar{36}\smallskip

\bcbar[value=,text={\emph{b.} There is a hidden determinism:}]{0}
\bcbar{33}\smallskip

\bcbar[value=,text={\emph{c.} The randomness is irreducible:}]{0}
\bcbar{26}\smallskip

\bcbar[value=,text={\emph{d.} Randomness is a fundamental concept in nature:}]{0}
\bcbar{24}

\bcxlabel{percent of votes}
\end{bchart}}

\noindent 15\%  made marginal comments, e.g., ``stupid answers'', ``don't know'', ``I'm undecided''.

\vspace{.3in}

\subsubsection*{Question 2: Do you believe that physical objects have their properties well defined prior to and independent of measurement?}

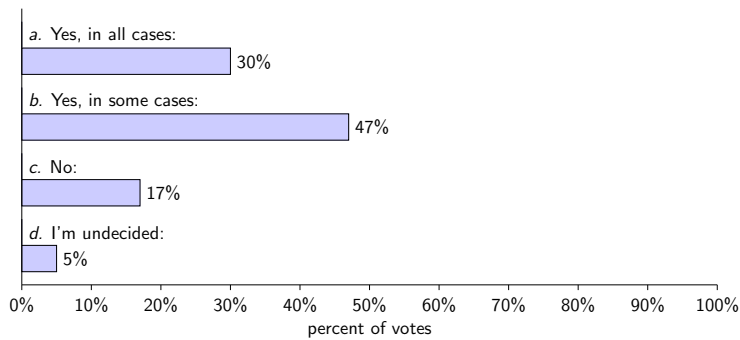
\scalebox{0.7}{\begin{bchart}[step=10,max=100,unit=\%,width=0.8\textwidth]

\bcbar[value=,text={\emph{a.} Yes, in all cases:}]{0}
\bcbar{30}\smallskip

\bcbar[value=,text={\emph{b.} Yes, in some cases:}]{0}
\bcbar{47}\smallskip

\bcbar[value=,text={\emph{c.} No:}]{0}
\bcbar{17}\smallskip

\bcbar[value=,text={\emph{d.} I'm undecided:}]{0}
\bcbar{5}

\bcxlabel{percent of votes}
\end{bchart}}

\noindent  7\% made marginal comments, e.g., ``formulation too unclear'', ``Give me a definition of `property'.''

\vspace{.3in}

\subsubsection*{Question 3: Einstein's view of quantum mechanics}

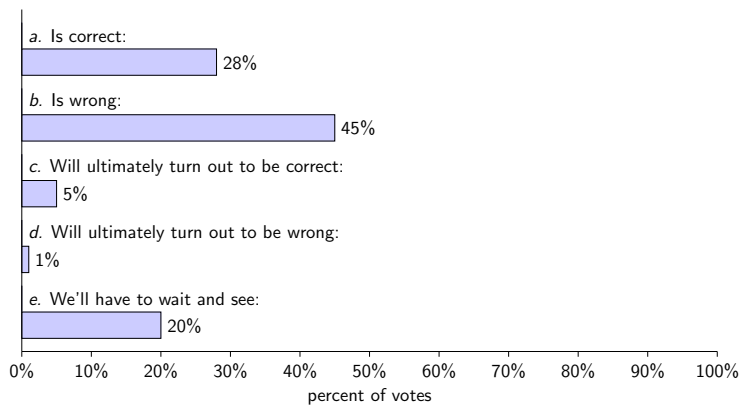
\scalebox{0.7}{\begin{bchart}[step=10,max=100,unit=\%,width=0.8\textwidth]

\bcbar[value=,text={\emph{a.} Is correct:}]{0}
\bcbar{28}\smallskip

\bcbar[value=,text={\emph{b.} Is wrong:}]{0}
\bcbar{45}\smallskip

\bcbar[value=,text={\emph{c.} Will ultimately turn out to be correct:}]{0}
\bcbar{5}\smallskip

\bcbar[value=,text={\emph{d.} Will ultimately turn out to be wrong:}]{0}
\bcbar{1}\smallskip

\bcbar[value=,text={\emph{e.} We'll have to wait and see:}]{0}
\bcbar{20}

\bcxlabel{percent of votes}
\end{bchart}}

\noindent 18\% made marginal comments, e.g., ``what the heck is meant exactly?'', ``can striving for a deeper understanding be considered correct or incorrect?'', ``partly right, partly wrong''.


\subsubsection*{Question 4: Bohr's view of quantum mechanics}

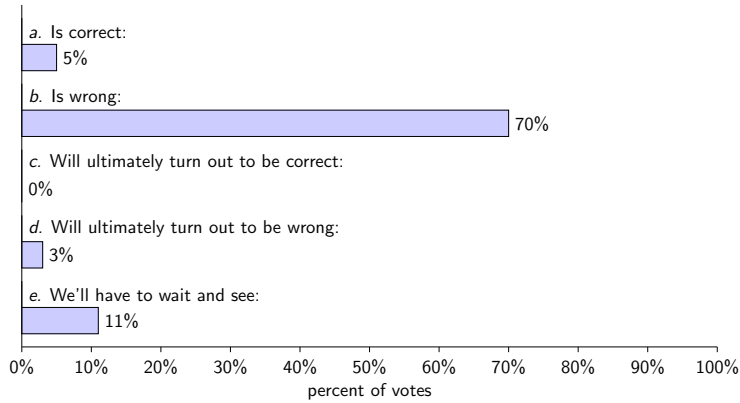
\scalebox{0.7}{\begin{bchart}[step=10,max=100,unit=\%,width=0.8\textwidth]

\bcbar[value=,text={\emph{a.} Is correct:}]{0}
\bcbar{5}\smallskip

\bcbar[value=,text={\emph{b.} Is wrong:}]{0}
\bcbar{70}\smallskip

\bcbar[value=,text={\emph{c.} Will ultimately turn out to be correct:}]{0}
\bcbar{0}\smallskip

\bcbar[value=,text={\emph{d.} Will ultimately turn out to be wrong:}]{0}
\bcbar{3}\smallskip

\bcbar[value=,text={\emph{e.} We'll have to wait and see:}]{0}
\bcbar{11}

\bcxlabel{percent of votes}
\end{bchart}}

\noindent  12\% made marginal comments, e.g., ``don't know'', ``none of the above'', ``I like some aspects of it, but there is some bad philosophy related to it'', ``partly fruitful, partly obstructive (regressive)''.

\vspace{.15in}

\subsubsection*{Question 5: The measurement problem}

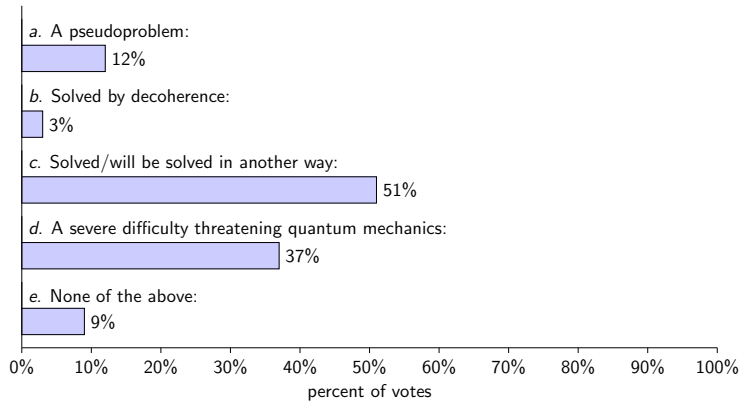
\scalebox{0.7}{\begin{bchart}[step=10,max=100,unit=\%,width=0.8\textwidth]

\bcbar[value=,text={\emph{a.} A pseudoproblem:}]{0}
\bcbar{12}\smallskip

\bcbar[value=,text={\emph{b.} Solved by decoherence:}]{0}
\bcbar{3}\smallskip

\bcbar[value=,text={\emph{c.} Solved/will be solved in another way:}]{0}
\bcbar{51}\smallskip

\bcbar[value=,text={\emph{d.} A severe difficulty threatening quantum mechanics:}]{0}
\bcbar{37}\smallskip

\bcbar[value=,text={\emph{e.} None of the above:}]{0}
\bcbar{9}

\bcxlabel{percent of votes}
\end{bchart}}

\noindent 8\% made marginal comments, e.g., ``solved with a conceptual clarification (see Primitive Ontology)''.

\vspace{.15in}

\subsubsection*{Question 6: What is the message of the observed violations of Bell's inequalities?}

\noindent \scalebox{0.7}
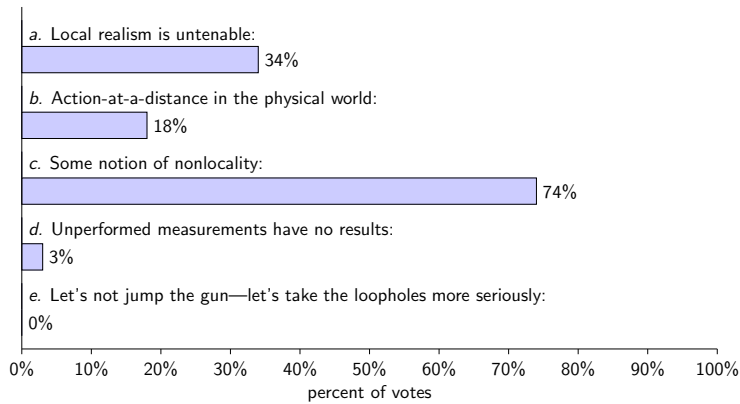
{\begin{bchart}[step=10,max=100,unit=\%,width=0.8\textwidth]

\bcbar[value=,text={\emph{a.} Local realism is untenable:}]{0}
\bcbar{34}\smallskip

\bcbar[value=,text={\emph{b.} Action-at-a-distance in the physical world:}]{0}
\bcbar{18}\smallskip

\bcbar[value=,text={\emph{c.} Some notion of nonlocality:}]{0}
\bcbar{74}\smallskip

\bcbar[value=,text={\emph{d.} Unperformed measurements have no results:}]{0}
\bcbar{3}\smallskip

\bcbar[value=,text={\emph{e.} Let's not jump the gun---let's take the loopholes more seriously:}]{0}
\bcbar{0}

\bcxlabel{percent of votes}
\end{bchart}}

\noindent 3\% made marginal comments, e.g., ``no strong opinion''.

\vspace{.3in}

\subsubsection*{Question 7: What about quantum information?}

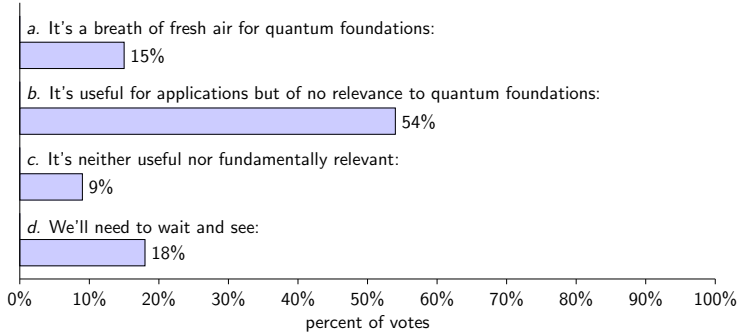
\scalebox{0.7}{\begin{bchart}[step=10,max=100,unit=\%,width=0.8\textwidth]

\bcbar[value=,text={\emph{a.} It's a breath of fresh air for quantum foundations:}]{0}
\bcbar{15}\smallskip

\bcbar[value=,text={\emph{b.} It's useful for applications but of no relevance to quantum foundations:}]{0}
\bcbar{54}\smallskip

\bcbar[value=,text={\emph{c.} It's neither useful nor fundamentally relevant:}]{0}
\bcbar{9}\smallskip

\bcbar[value=,text={\emph{d.} We'll need to wait and see:}]{0}
\bcbar{18}

\bcxlabel{percent of votes}
\end{bchart}}

\noindent 5\% made marginal comments, e.g.,  ``don't know'', ``(e) some relevance to quantum foundations''.

\vspace{.3in}

\subsubsection*{Question 8: When will we have a working and useful quantum computer?}

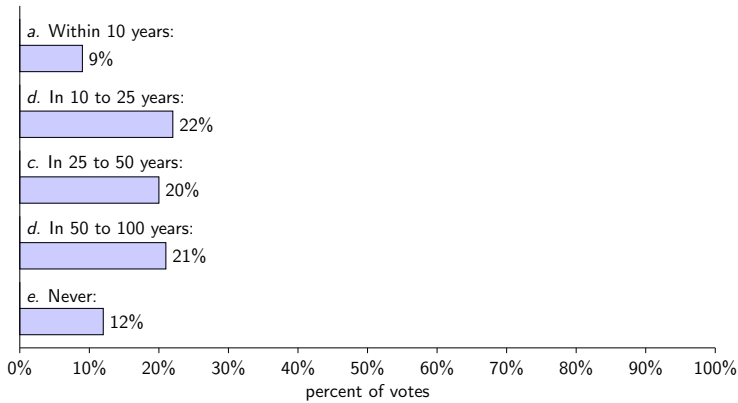
\scalebox{0.7}{\begin{bchart}[step=10,max=100,unit=\%,width=0.8\textwidth]

\bcbar[value=,text={\emph{a.} Within 10 years:}]{0}
\bcbar{9}\smallskip

\bcbar[value=,text={\emph{d.} In 10 to 25 years:}]{0}
\bcbar{22}\smallskip

\bcbar[value=,text={\emph{c.} In 25 to 50 years:}]{0}
\bcbar{20}\smallskip

\bcbar[value=,text={\emph{d.} In 50 to 100 years:}]{0}
\bcbar{21}\smallskip

\bcbar[value=,text={\emph{e.} Never:}]{0}
\bcbar{12}

\bcxlabel{percent of votes}
\end{bchart}}

\noindent 13\% made marginal comments, e.g.,  ``?'', ``don't know'', ``what is `useful'?''

\vspace{.3in}

\subsubsection*{Question 9: What interpretation of quantum states do you prefer?}

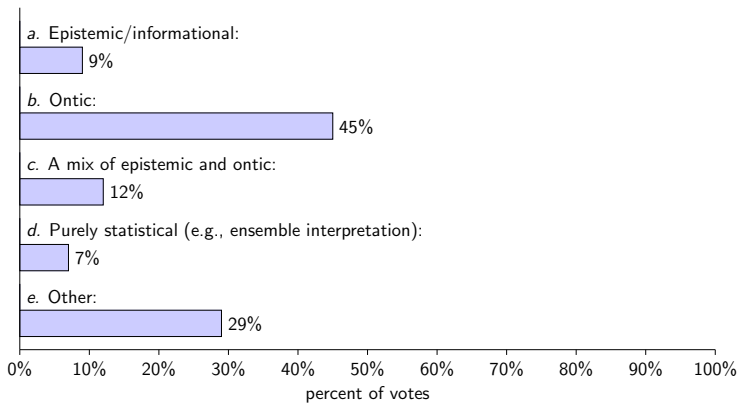
\scalebox{0.7}{\begin{bchart}[step=10,max=100,unit=\%,width=0.8\textwidth]

\bcbar[value=,text={\emph{a.} Epistemic/informational:}]{0}
\bcbar{9}\smallskip

\bcbar[value=,text={\emph{b.} Ontic:}]{0}
\bcbar{45}\smallskip

\bcbar[value=,text={\emph{c.} A mix of epistemic and ontic:}]{0}
\bcbar{12}\smallskip

\bcbar[value=,text={\emph{d.} Purely statistical (e.g., ensemble interpretation):}]{0}
\bcbar{7}\smallskip

\bcbar[value=,text={\emph{e.} Other:}]{0}
\bcbar{29}

\bcxlabel{percent of votes}
\end{bchart}}

\noindent 5\% made marginal comments, e.g., ``stupid question'', ``no strong opinion'', ``define quantum state''.

\vspace{.3in}

\subsubsection*{Question 10: The observer}

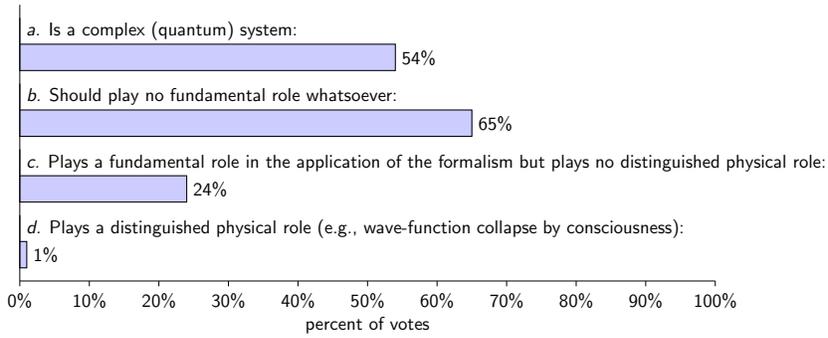
\scalebox{0.7}{\begin{bchart}[step=10,max=100,unit=\%,width=0.8\textwidth]

\bcbar[value=,text={\emph{a.} Is a complex (quantum) system:}]{0}
\bcbar{54}\smallskip

\bcbar[value=,text={\emph{b.} Should play no fundamental role whatsoever:}]{0}
\bcbar{65}\smallskip

\bcbar[value=,text={\emph{c.} Plays a fundamental role in the application of the formalism but plays no distinguished physical role:}]{0}
\bcbar{24}\smallskip

\bcbar[value=,text={\emph{d.} Plays a distinguished physical role (e.g., wave-function collapse by consciousness):}]{0}
\bcbar{1}

\bcxlabel{percent of votes}
\end{bchart}}

\noindent 8\% made marginal comments, e.g., ``none of the above''.

\vspace{2in}

\subsubsection*{Question 11: Reconstructions of quantum theory}

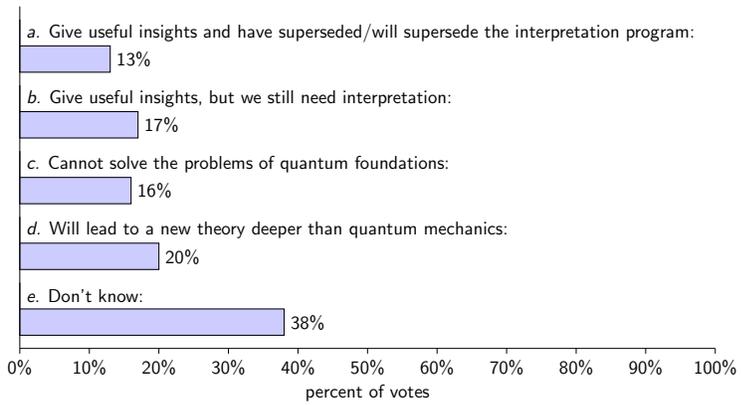
\scalebox{0.7}{\begin{bchart}[step=10,max=100,unit=\%,width=0.8\textwidth]

\bcbar[value=,text={\emph{a.} Give useful insights and have superseded/will supersede the interpretation program:}]{0}
\bcbar{13}\smallskip

\bcbar[value=,text={\emph{b.} Give useful insights, but we still need interpretation:}]{0}
\bcbar{17}\smallskip

\bcbar[value=,text={\emph{c.} Cannot solve the problems of quantum foundations:}]{0}
\bcbar{16}\smallskip

\bcbar[value=,text={\emph{d.} Will lead to a new theory deeper than quantum mechanics:}]{0}
\bcbar{20}\smallskip

\bcbar[value=,text={\emph{e.} Don't know:}]{0}
\bcbar{38}

\bcxlabel{percent of votes}
\end{bchart}}

\noindent 9\%  made marginal comments, e.g., ``??''

\vspace{.1in}

\subsubsection*{Question 12: What is your favorite interpretation of quantum mechanics?}

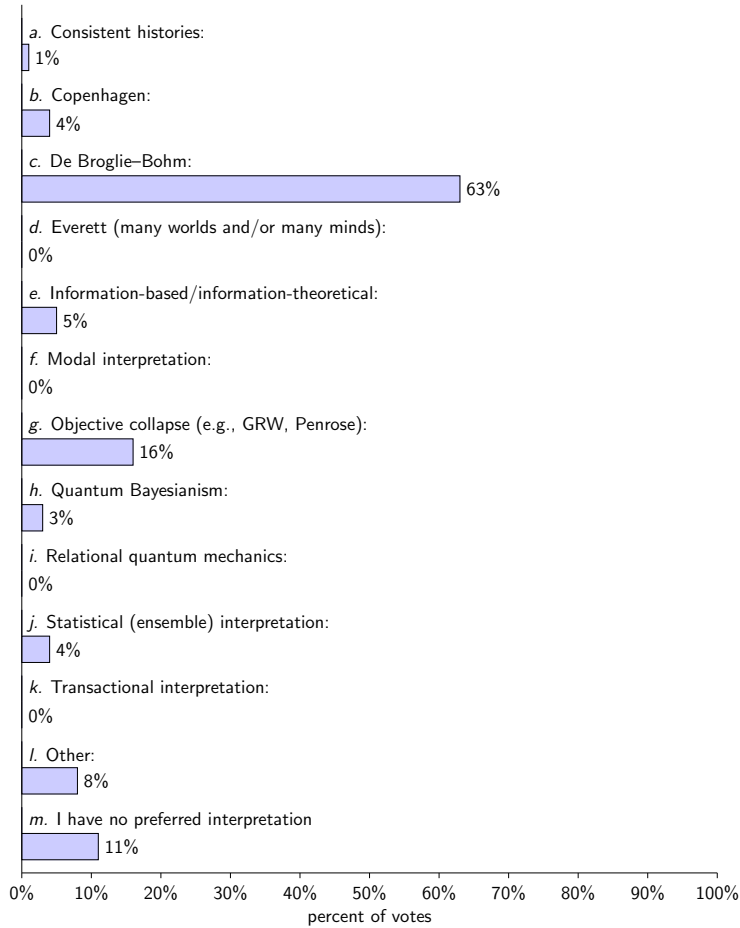
\scalebox{0.7}{\begin{bchart}[step=10,max=100,unit=\%,width=0.8\textwidth]

\bcbar[value=,text={\emph{a.} Consistent histories:}]{0}
\bcbar{1}\smallskip

\bcbar[value=,text={\emph{b.} Copenhagen:}]{0}
\bcbar{4}\smallskip

\bcbar[value=,text={\emph{c.} De Broglie--Bohm:}]{0}
\bcbar{63}\smallskip

\bcbar[value=,text={\emph{d.} Everett (many worlds and/or many minds):}]{0}
\bcbar{0}\smallskip

\bcbar[value=,text={\emph{e.} Information-based/information-theoretical:}]{0}
\bcbar{5}\smallskip

\bcbar[value=,text={\emph{f.} Modal interpretation:}]{0}
\bcbar{0}\smallskip

\bcbar[value=,text={\emph{g.} Objective collapse (e.g., GRW, Penrose):}]{0}
\bcbar{16}\smallskip

\bcbar[value=,text={\emph{h.} Quantum Bayesianism:}]{0}
\bcbar{3}\smallskip

\bcbar[value=,text={\emph{i.} Relational quantum mechanics:}]{0}
\bcbar{0}\smallskip

\bcbar[value=,text={\emph{j.} Statistical (ensemble) interpretation:}]{0}
\bcbar{4}\smallskip

\bcbar[value=,text={\emph{k.} Transactional interpretation:}]{0}
\bcbar{0}\smallskip

\bcbar[value=,text={\emph{l.} Other:}]{0}
\bcbar{8}\smallskip

\bcbar[value=,text={\emph{m.} I have no preferred interpretation}]{0}
\bcbar{11}

\bcxlabel{percent of votes}
\end{bchart}}

\noindent 4\% made marginal comments, e.g., ``iff [option c] means Bohmian Mechanics''.

\vspace{.3in}

\subsubsection*{Question 13: How often have you switched to a different interpretation?}

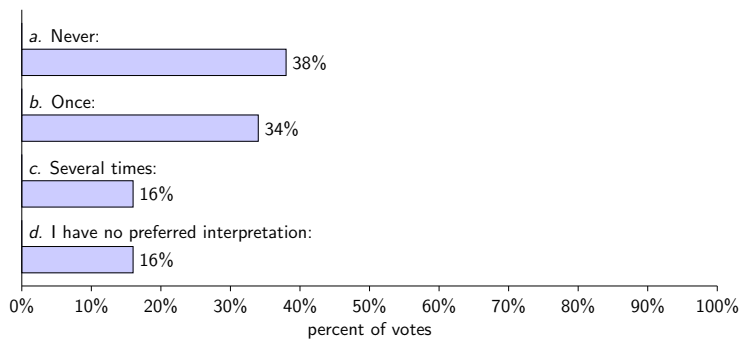
\scalebox{0.7}{\begin{bchart}[step=10,max=100,unit=\%,width=0.8\textwidth]

\bcbar[value=,text={\emph{a.} Never:}]{0}
\bcbar{38}\smallskip

\bcbar[value=,text={\emph{b.} Once:}]{0}
\bcbar{34}\smallskip

\bcbar[value=,text={\emph{c.} Several times:}]{0}
\bcbar{16}\smallskip

\bcbar[value=,text={\emph{d.} I have no preferred interpretation:}]{0}
\bcbar{16}

\bcxlabel{percent of votes}
\end{bchart}}

\noindent 5\% made marginal comments, e.g., ``can't decide if switching from `not thinking about it' to `Bohm' should count as switching to a different interpretation'', ``before I was only confused'', ``I'm just trying to improve my understanding of quantum physics. It's good to have all the consistent theories (interpretations) on the table.''

\vspace{.3in}

\subsubsection*{Question 14: How much is the choice of interpretation a matter of personal philosophical prejudice?}

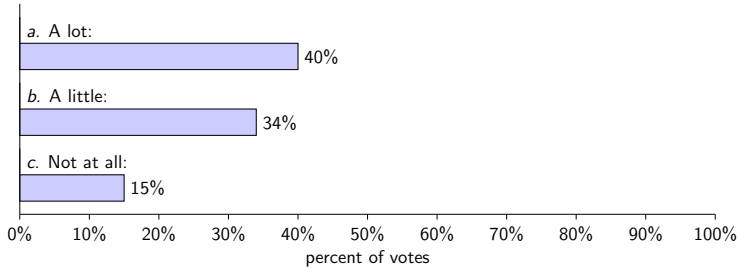
\scalebox{0.7}{\begin{bchart}[step=10,max=100,unit=\%,width=0.8\textwidth]

\bcbar[value=,text={\emph{a.} A lot:}]{0}
\bcbar{40}\smallskip

\bcbar[value=,text={\emph{b.} A little:}]{0}
\bcbar{34}\smallskip

\bcbar[value=,text={\emph{c.} Not at all:}]{0}
\bcbar{15}

\bcxlabel{percent of votes}
\end{bchart}}

\noindent 7\% made marginal comments, e.g., ``stupid question'', ``There should be no need for interpretation!''

\vspace{.3in}

\subsubsection*{Question 15: Superpositions of macroscopically distinct states}

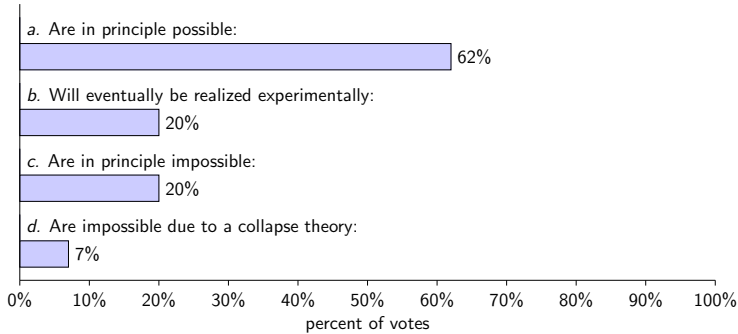
\scalebox{0.7}{\begin{bchart}[step=10,max=100,unit=\%,width=0.8\textwidth]

\bcbar[value=,text={\emph{a.} Are in principle possible:}]{0}
\bcbar{62}\smallskip

\bcbar[value=,text={\emph{b.} Will eventually be realized experimentally:}]{0}
\bcbar{20}\smallskip

\bcbar[value=,text={\emph{c.} Are in principle impossible:}]{0}
\bcbar{20}\smallskip

\bcbar[value=,text={\emph{d.} Are impossible due to a collapse theory:}]{0}
\bcbar{7}

\bcxlabel{percent of votes}
\end{bchart}}

\noindent 8\% made marginal comments, e.g., ``?''

\vspace{.3in}

\subsubsection*{Question 16: In 50 years, will we still have conferences devoted to quantum foundations?}

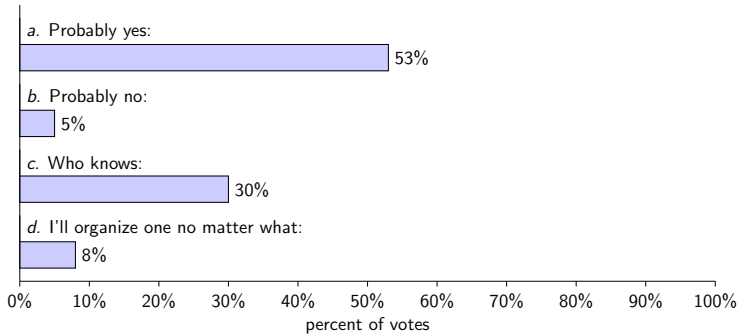
\scalebox{0.7}{\begin{bchart}[step=10,max=100,unit=\%,width=0.8\textwidth]

\bcbar[value=,text={\emph{a.} Probably yes:}]{0}
\bcbar{53}\smallskip

\bcbar[value=,text={\emph{b.} Probably no:}]{0}
\bcbar{5}\smallskip

\bcbar[value=,text={\emph{c.} Who knows:}]{0}
\bcbar{30}\smallskip

\bcbar[value=,text={\emph{d.} I'll organize one no matter what:}]{0}
\bcbar{8}

\bcxlabel{percent of votes}
\end{bchart}}

\noindent 9\% made marginal comments, e.g.,  ``who is `we'?'', ``I hope not'', ``I hope they aren't necessary anymore!'', ``no, we will ask the quantum computer (see Q8) and it will have inherent understanding of what a quantum state is, and will explain it to us''.

\newpage

\section{\label{sec:discussion}Discussion}

In many ways the results of the survey speak for themselves.  Of particular interest, though, are the several ways in which our picture differs substantially from the snapshot taken by SKZ.  To quantify this, we computed the square of the difference $d$ in response rate for each given option, and then summed this over all given options for each question.  The three questions with the highest $\Sigma d^2$ were, in decreasing order:  Q12, Q7, and Q6.  We discuss each of these briefly:

\vspace{.2in}

\noindent {\bf{Question 12:  What is your favorite interpretation of
    quantum mechanics?}}  In the SKZ results, \emph{b. Copenhagen}
(42\%) and \emph{e. Information-based/information-theoretical} (24\%)
received the highest response rates, while \emph{c. de Broglie - Bohm}
received zero votes of endorsement.  SKZ write explicitly that ``the
fact that de Broglie - Bohm interpretation did not receive any votes
may simply be an artifact of the particular set of participants we
polled.''  Our results strongly confirm this suspicion.  At the
Bielefeld conference, choice \emph{c. de Broglie - Bohm} garnered far
and away the majority of the votes (63\%) while \emph{b. Copenhagen}
and \emph{e. information-based / information-theoretical} received a
paltry 4\% and 5\% respectively.  It is also interesting to compare
results on this question to the older (1997) survey conducted by Max
Tegmark.  \cite{tegmark}  Tegmark, finding that 17\% of his
respondents endorsed a many-worlds / Everett interpretation, announced
this as a ``rather striking shift in opinion compared to the old days
when the Copenhagen interpretation reigned supreme.''  Our results
clearly suggest, though, that any such interpretation of these sorts
of poll results -- as indicating a meaningful temporal shift in
attitudes -- should be taken with a rather large grain of salt.  It is
almost certainly not the case, for example, that while a ``striking
shift'' toward many-worlds views occured in the years prior to 1997,
this shift then stalled out between 1997 and 2011 (the response rate
endorsing Everett being about the same in the Tegmark and SKZ polls),
and then suddenly collapsed (with the majority of quantum foundations
workers now embracing the de Broglie - Bohm pilot-wave theory).
Instead, the obviously more plausible interpretation of the data is
that each poll was given to a very different and highly
non-representative group.  The snapshots reveal much more about the
processes by which it was decided whom should be invited to a given
conference, than they reveal about trends in the thinking of the
community as a whole.  We note finally that insofar as our poll got
more than twice as many respondents as the SKZ poll (which those
authors had described as ``the most comprehensive poll of
quantum-foundational views ever conducted'') it is now apparently the
case that the de Broglie - Bohm pilot-wave theory is, by an incredibly
large margin, the most endorsed interpretation in the most
comprehensive poll of quantum-foundational views ever conducted.  For
the reasons we have just been explaining, this has almost no meaning,
significance, or implications, beyond the fact that lots of
``Bohmians'' were invited to the Bielefeld conference.  But it does
demonstrate rather strikingly that the earlier conferences (where
polls were conducted by Tegmark and SKZ) somehow failed to involve a
rather large contingent of the broader foundations community.  And
similarly, the Bielefeld conference somehow failed to involve the
large Everett-supporting contingent of the broader foundations community.

\vspace{.2in}

\noindent{\bf{Question 7:  What about quantum information?}}  In the
SKZ poll, \emph{a. It's a breath of fresh air for quantum foundations}
received an overwhelming majority of votes (76\%) and was indeed the
most-endorsed answer on the entire poll; a mere 6\% of respondents
selected \emph{b. It's useful for applications but of no relevance to
  quantum foundations}.  In our poll the situation was reversed:  only
15\% of respondents thought quantum information was a breath of fresh
air, while 54\% thought it was useful for applications but of no
relevance to quantum foundations.   This dramatic difference again
almost certainly does not signal a seismic shift in the field during
the year or so between the two polls, but instead arises from the
apparently large difference between the two populations polled.  In
particular, it is not terribly surprising that endorsing the de
Broglie - Bohm interpretation correlates positively with endorsing
\emph{b.} on this question, and that endorsing either the Copenhagen
or information-based interpretations correlates with answering
\emph{a.} here.  Most people who like the de Broglie - Bohm theory do
so precisely because it provides a candidate account of quantum
phenomena in which no reference to anthropocentric notions (like
``measurement'', ``observation'', ``information'', etc.) need appear
in the formulation of the theory.  This, indeed, was the theme of the
Bielefeld conference, literally stated in its title:  ``Quantum Theory
Without Observers''.  So it is no surprise that attendees at this
particular conference would tend to have more ``realist'' outlooks,
would tend to be attracted to theories like de Broglie - Bohm, and
would tend to think that quantum information (whatever its merits and
virtues for applications) is not appropriate as an irreducible foundation for
understanding the physics of quantum phenomena.

\vspace{.2in}

\noindent {\bf{Question 6:  What is the message of the observed
    violations of Bell's inequalities?}}  In the SKZ poll, two
different responses -- \emph{a. Local realism is untenable} and
\emph{d. Unperformed measurements have no results} -- both received
more than 50\% endorsement.  (Recall that multiple responses were
allowed!)  Evidently then a large fraction of SKZ's respondents believe
that both \emph{a.} and \emph{d.} can be concluded from Bell's theorem
and the associated experimental tests.  Presumably this reflects the
belief that several assumptions -- ``locality'', ``realism'', and the
idea that ``unperformed measurements \emph{do} have results'' -- are
needed for the derivation of the empirically-excluded Bell
inequalities.   In our poll, on the other
hand, hardly anyone (a mere 3\% of respondents) endorsed \emph{d.}  A
huge majority (74\%) selected \emph{c. Some notion of nonlocality},
and the related/overlapping answers \emph{b. Action-at-a-distance
  in the physical world} and \emph{a.  Local realism is untenable}
also received significant support (18\% and 34\% respectively).  Note
that anybody who believes that the observed violations of Bell's
inequalities implies ``some notion of nonlocality'' \emph{ipso facto}
must also believe that ``local realism is untenable'', so the
significant overlap there makes sense.  Evidently, then, most
respondents at the Bielefeld conference believe that only one
assumption -- ``locality'' -- is needed for the derivation of the
empirically-excluded Bell inequalities.   In our opinion, it is here on
this question that the difference in the responses between the two polls
is most interesting and most surprising (or at least should appear
most surprising to someone from outside the foundations community).  
Whereas Questions 12 and 7 ask respondents to assess the merits
of a certain interpretation or viewpoint or research program, this
Question 6 is essentially asking:  what are the premises of a certain
mathematical theorem?  It is somehow not terribly surprising that
different sub-groups within the foundations community would have
different background assumptions that lead them to judge different
interpretations/programs as ``scientifically the best'' or ``most
likely to lead to important future progress'' or whatever.  But it
\emph{is} terribly surprising that different sub-groups could
continue, after all these decades, to disagree about what minimal set
of assumptions is needed to derive the Bell-type inequalities that
(almost) everyone agrees are incompatible with experimental data.  
If there is an ``embarrassment'' \cite{plumer}  to be found in any of the poll results,
it lies here.

\vspace{.2in}

There are several other significant and interesting differences between the results of
the two polls, but the above three stand out not only quantitatively
but also in terms of the clear centrality and fundamentality of the issues
involved.   Probably the only thing that can be inferred with
statistical confidence from the results is that on these several
fundamental questions, the two polled groups were quite different.
What explains this?  The answer is obvious and has already been
alluded to above:  each group consisted of a
special sub-set of researchers in quantum foundations ... special
in that they had been invited to the conference in question.  

Several of the organizers of the Bielefeld conference, for example,
are prominent proponents of the de Broglie - Bohm pilot-wave theory.  And in
general the ``Quantum Theory Without Observers'' series of conferences
has been dedicated to furthering the work and ideas of John Bell,
whose image for example graced the conference poster.  \cite{QTWO3}
It is thus not terribly surprising that people invited to attend this
conference were -- by no means exclusively, but, compared to the
community at large, unusually -- sympathetic to the views developed
and endorsed by Bell, including ``realism'' (meaning here the inappropriateness of
anthropocentric concepts like ``information'' or ``measurement'' appearing in the
formulation of fundamental theories), an extremely high regard for the
de Broglie - Bohm pilot-wave theory, and an insistence that it is
\emph{locality} (and not some other notion such as ``realism'' or
``determinism'') that is called into question by the experimental
tests of Bell's inequalities. \cite{bell}  From this point of view,
the results of our poll are hardly shocking:  they indicate
only that people who are both motivated to and invited to participate
in a workshop largely celebrating Bell's continuing influence on the
foundations of quantum theory, tend to answer questions similarly to
how Bell himself would have answered them. 

The different answers given by respondents in the SKZ poll can perhaps
be understood similarly -- for example, by reference to the fact that
Anton Zeilinger, one of the leading proponents of a kind of
neo-Copenhagen approach to understanding quantum theory, was one of
the organizers \cite{gg} or perhaps by reference to the fact that the invitees
to the conference at which SKZ's poll was given out were people given
a certain kind of grant during a certain period of time by the
Templeton Foundation.  \cite{pc}
Presumably the results of another recent survey which used the same set of
questions \cite{sommer} could be understood in a similar way, namely,
as telling one more about the biases inherent in the invitation
process than it does about what experts in quantum foundations
generally think.  

With the possible exception of the case discussed in Ref.~\cite{gg},  
it is not at all our goal to criticize the existence of bias
in the determination of whom should be invited to attend a given
conference.  ``Bias'' here simply means that potential attendees are
selected in accordance with the extent to which their individual interests
and perspectives align with the goals and themes of the conference,
and it is entirely reasonable and proper for conference organizers to
choose such goals and themes and indeed to aim for a healthy but
non-disruptive representation by opposing views.  
Our point in stressing the role of attendance-bias is instead this:
none of these polls, our own very much included, should be taken too
seriously as capturing a meaningful snapshot of anything but the views
of a small and biased minority.

\subsubsection*{Acknowledgements}
Thanks to the organizers of Quantum Theory Without Observers III for
letting us inflict the survey on the participants, and thanks to the
participants who so graciously took the time to fill it out.


\begin{thebibliography}{1}

\bibitem{SKZ}
M. Schlosshauer, J. Kofler, and A. Zeilinger, ``A Snapshot of Foundational Attitudes Toward Quantum Mechanics'', arxiv/1301.1069

\bibitem{ball} P. Ball, ``Experts still split about what quantum
  theory means'', \emph{Nature},  January 11, 2013

\bibitem{moskowitz} C. Moskowitz, ``Physicists disagree over meaning
  of quantum mechanics, poll shows'', January 21, 2013; \url{http://nbcnews.com}

\bibitem{plumer} B. Plumer, ``Why quantum mechanics is an
  `embarrassment' to science'', February 7, 2013; \url{http://www.washingtonpost.com/blogs/wonkblog}


\bibitem{mit}
``Poll reveals quantum physicists' disagreement about the nature of
reality'', MIT Technology Review, January 9, 2013; \url{http://m.technologyreview.com}

\bibitem{siegfried}
T. Siegfried, ``Poll of quantum physicists shows agreement,
disagreement and something in between'', \emph{Science News}, 
February 20, 2013; \url{http://www.sciencenews.org}


\bibitem{schloss}
M. Schlosshauer, ``Agreeing to disagree'',
\emph{Physics World}, March 2013

\bibitem{QTWO3}
The conference website, which includes a list of participants, is here:
\url{http://www.mathematik.uni-muenchen.de/~bohmmech/bielefeld/index.html}

\bibitem{bell}  For Bell's views on the role of ``information'',
  ``measurement'', etc., in fundamental physical theories, see his
  paper ``Against `Measurement'.''  Almost all of Bell's papers on the
  foundations of quantum theory involve the de Broglie - Bohm
  pilot-wave theory in some way; see especially ``Quantum Mechanics
  for Cosmologists'' and ``On the impossible pilot wave''.    For
  Bell's view on what, exactly, is at stake in regard to the
  experimental tests of his famous inequality, see especially
  ``Bertlmann's socks and the nature of reality'' (including the
  illuminating footnote 10) and ``La nouvelle cuisine''.  All of these
  papers are conveniently reprinted in J.S. Bell, \emph{Speakable and
    Unspeakable in Quantum Mechanics}, Second Edition, Cambridge
  University Press, 2004. 


\bibitem{tegmark}
M. Tegmark, \emph{Fortschr. Phys.} {\bf{46}}, 855 (1998); arxiv:quant-ph/9709032

\bibitem{gg} Interestingly, there seems to be a historical precedent
  for  Zeilinger's failure to invite, to a relevant conference, people
  whose views overlap with those of John Bell:  see Gerhard Gr\"ossing, ``Serious Matter: The John Bell
  Scandal'', \url{http://www.nonlinearstudies.at/Bell_E.php}

\bibitem{pc}  See here for the relevant announcement:
  \url{http://www.templeton.org/what-we-fund/} \\ \url{funding-priorities/quantum-physics-and-the-nature-of-reality}


\bibitem{sommer} C. Sommer, ``Another Survey of Foundational Attitudes
  Towards Quantum Mechanics'', arxiv:1303.2719

\end{thebibliography}
\end{document}